\documentclass[reprint,amsmath,amssymb,aps,prd,showpacs]{revtex4-1}

\usepackage{graphicx}
\usepackage{dcolumn}
\usepackage{bm}
\usepackage{epsfig}
\usepackage{subfigure}
\usepackage{multirow}
\usepackage{amsmath} 
\usepackage{hyperref}

\begin{document}

\title{Implications of texture zeros for a variant of tribimaximal mixing}

\author{Sanjeev Kumar}\email{skverma@physics.du.ac.in}
 \affiliation{Department of Physics and Astrophysics, University of Delhi,\\ 
Delhi -110007, INDIA.} 

\author{Radha Raman Gautam}\email{gautamrrg@gmail.com}
 \affiliation{Department of Physics, Himachal Pradesh University, Shimla -171005, INDIA.} 

\date{\today}

\begin{abstract}
We study the phenomenological implications of the presence of two texture zeros 
in the neutrino mass matrix assuming that the neutrino mixing matrix 
has its first column identical to that of the tribimaximal mixing matrix.
Only two patterns of this kind are compatible with the experimental data.
These textures have definite predictions for the neutrino observables
that are testable in future neutrino experiments. 
\end{abstract}


\maketitle

In the recent years, considerable efforts have been put towards
determining the structure of the neutrino mass matrix ($M_\nu$)\cite{reviews}.
The non-zero value of the reactor mixing angle $\theta_{13}$,
recently determined in various neutrino oscillation
experiments \cite{th13}, has called many neutrino mass models
predicting $\theta_{13} = 0$ into question.
The models based upon the Tribimaximal (TBM) \cite{hps} mixing,
that predicted the reactor, atmospheric and solar mixing angles as
$\theta_{13} = 0$, $\theta_{23} = \frac{\pi}{2}$,
and $\theta_{12}=\arctan(1/ \sqrt{2})$, respectively, need modifications
in the light of a non-zero $\theta_{13}$. 
The TBM mixing matrix is given as
\begin{equation}
U_{\text{TBM}}=
\left(
\begin{array}{ccc}
 \sqrt{\frac{2}{3}} & \frac{1}{\sqrt{3}} & 0 \\
 -\frac{1}{\sqrt{6}} & \frac{1}{\sqrt{3}} &
   \frac{1}{\sqrt{2}} \\
 -\frac{1}{\sqrt{6}} & \frac{1}{\sqrt{3}} &
   -\frac{1}{\sqrt{2}} \\
\end{array}
\right).
\end{equation}

Many ways have been proposed to modify the TBM ansatz to
accommodate a non-zero $\theta_{13}$. A simple possibility is to  keep
one column of TBM mixing matrix unchanged
while modifying its other two columns within unitarity constraints \cite{partial}.
This gives rise to three Trimaximal (TM) mixing patterns, viz., TM$_1$,
TM$_2$, and TM$_3$, that have their first, second and third columns
identical to TBM matrix, respectively \cite{partial}. These three mixing schemes
contain TBM mixing as a special case that enlarges the symmetry of
these mixing patterns. Hence, they could have been named TBM$_1$, TBM$_2$,
and TBM$_3$.

TM$_1$ mixing is given as
\begin{equation}\label{eq:tm1}
U_{\text{TM}_1}=\left(
\begin{array}{ccc}
  \frac{2}{\sqrt{3}} & \frac{1}{\sqrt{3}} \cos \theta &
\frac{1}{\sqrt{3}} \sin \theta \\
  -\frac{1}{\sqrt{6}} &
      \frac{1}{\sqrt{3}}\cos\theta-\frac{e^{i \phi} \sin
\theta}{\sqrt{2}} &
\frac{1}{\sqrt{3}}\sin\theta+
\frac{e^{i \phi} \cos\theta}{\sqrt{2}} \\
  -\frac{1}{\sqrt{6}} &
 \frac{1}{\sqrt{3}}\cos\theta+\frac{e^{i \phi}
                        \sin \theta}{\sqrt{2}} &
 \frac{1}{\sqrt{3}}\sin \theta
-\frac{e^{i \phi}
   \cos \theta}{\sqrt{2}}\end{array}
\right).
\end{equation}
The mixing scheme reduces to the TBM scheme in the special case 
$\theta=0$ and $\phi=0$. 

TM$_2$ mixing is given as
\begin{equation}\label{eq:tm2}
U_{\text{TM}_2}=\left(
\begin{array}{ccc}
 \sqrt{\frac{2}{3}} \cos \theta &
   \frac{1}{\sqrt{3}} & \sqrt{\frac{2}{3}}
   \sin \theta \\
 -\frac{\cos\theta}{\sqrt{6}}+\frac{e^{-i \phi} \sin
\theta}{\sqrt{2}} & \frac{1}{\sqrt{3}} &
   -\frac{\sin\theta}{\sqrt{6}}-\frac{e^{-i \phi} \cos\theta}{\sqrt{2}} \\
 -\frac{\cos\theta}{\sqrt{6}}-\frac{e^{-i \phi}
   \sin \theta}{\sqrt{2}} &
   \frac{1}{\sqrt{3}} & -\frac{\sin
   \theta}{\sqrt{6}}
+\frac{e^{-i \phi}
   \cos \theta}{\sqrt{2}}\end{array}
\right).
\end{equation}
The mixing scheme reduces to the TBM scheme in the special case 
$\theta=0$ and $\phi=0$. This mixing scheme corresponds to the
magic symmetry.

TM$_3$ mixing is given as
\begin{equation}\label{eq:tm3}
U_{\text{TM}_3}=
\left(
\begin{array}{ccc}
 \cos \theta & \sin \theta & 0 \\
 -\frac{e^{-i \phi } \sin \theta}{\sqrt{2}} &
   \frac{e^{-i \phi } \cos \theta}{\sqrt{2}} &
   \frac{1}{\sqrt{2}} \\
 -\frac{e^{-i \phi } \sin \theta}{\sqrt{2}} &
   \frac{e^{-i \phi } \cos \theta}{\sqrt{2}} &
   -\frac{1}{\sqrt{2}}
\end{array}
\right).
\end{equation}
The mixing scheme reduces to the TBM scheme in the special case 
$\theta=\arctan (1/\sqrt{2})$ and $\phi=0$. This mixing scheme
is equivalent to $\mu-\tau$ symmetry.

Another simple assumption that can accommodate a non-zero $\theta_{13}$ is the presence of texture zeros in the neutrino
mass matrix \cite{fgm, xingtz, tz}. Texture zeros induce relations
between mixing matrix elements and neutrino masses. Considering neutrinos
to be Majorana fermions and working in a basis where the charged 
lepton mass matrix $M_l$ is diagonal, there are in total
fifteen different patterns of two texture zeros in the  
neutrino mass matrices. Out of these fifteen possible patterns,
only seven can satisfy the present neutrino oscillation
\mbox{data \cite{fgm, xingtz}}. These seven patterns are classified in three classes A, B and C corresponding to the normal, quasi-degenerate
and inverted mass hierarchies of neutrinos [Table \ref{tab:2t}].

\begin{table}[tb]
\begin{center}
\begin{tabular}{cc}
\hline
\hline
 Type  &        Constraining Equations         \\
 \hline
 A$_1$ &     $M_{ee}=0$, $M_{e\mu}=0$     \\
 A$_2$ &    $M_{ee}=0$, $M_{e\tau}=0$     \\
 B$_1$ &  $M_{e\tau}=0$, $M_{\mu\mu}=0$   \\
 B$_2$ &  $M_{e\mu}=0$, $M_{\tau\tau}=0$  \\
 B$_3$ &   $M_{e\mu}=0$, $M_{\mu\mu}=0$   \\
 B$_4$ & $M_{e\tau}=0$, $M_{\tau\tau}=0$  \\
 C   & $M_{\mu\mu}=0$, $M_{\tau\tau}=0$  \\
 \hline
 \hline
\end{tabular}
\end{center}
\caption{Seven allowed mass matrices with two zeros classified 
into three classes.}
\label{tab:2t}
\end{table}

The current neutrino data are consistent with the possibility of keeping
first or second column of the mixing matrix unmodified (TM$_1$ or TM$_2$
mixing) while modifying other columns within unitarity constraints.
The experimental data is also consistent with the presence of two texture zeros
in the neutrino mass matrix. If we combine both approaches together by
having texture zeros in a mass matrix corresponding to TM$_1$ or TM$_2$
mixing, we are bound to get very predictive neutrino mass matrices.
In an earlier work \cite{2tztm}, we studied the implications of two
texture zeros in a magic mass matrix giving TM$_2$ mixing.
In the present work, we study the phenomenological implications of the
presence of two texture zeros in a neutrino mass matrix giving TM$_1$ mixing.

A mass matrix giving TM$_1$ mixing can be parameterized as
\begin{equation}\label{eq:mtm1}
M_{\text{TM}_1} =\left(
\begin{array}{ccc}
 2 a & 2 b &2 c \\
 2 b & b-c+d & 2 a+2 b-d \\
 2 c & 2a+2b-d & -3b+3c+d
\end{array}
\right).
\end{equation}
We can obtain the form of the neutrino mass matrix giving TM$_1$ mixing
for the seven patterns of two texture zeros by substituting the respective
constraints from Table \ref{tab:2t} in Eq. (\ref{eq:mtm1}).

The neutrino mass matrix of type A$_1$ giving TM$_1$ mixing is 
given as
\begin{equation}\label{eq:a1}
 M^{\text{A}_1}= \left(
\begin{array}{ccc}
0 & 0 & 2 c \\ 0 & -2c+\Delta & c-\Delta\\ 2c & c-\Delta & 2c+\Delta 
\end{array}
\right)
\end{equation}
where $\Delta=d+b$. Similarly, the neutrino mass matrix of type A$_2$
giving TM$_1$ mixing is
\begin{equation}\label{eq:a2}
 M^{\text{A}_2} = \left(
\begin{array}{ccc}
0 & 2b & 0 \\ 2b & 2b+\Delta & b-\Delta \\ 0& b-\Delta & -2b+\Delta
\end{array}
\right),
\end{equation}
where $\Delta=d-c$.

The four mass matrices giving TM$_1$ mixing for class B are 
\begin{equation}
M^{\text{B}_1} =\left(
\begin{array}{ccc}
 2a & 2b & 0 \\
 2b & 0 & 2a +3b\\
 0 & 2a +3b& -4b
\end{array}
\right),
\end{equation}
\begin{equation}
M^{\text{B}_2} =\left(
\begin{array}{ccc}
 2a & 0 & 2c \\
 0 & -4c & 2a+3c \\
 2c & 2a+3c & 0
\end{array}
\right),
\end{equation}
\begin{equation}
M^{\text{B}_3} =\left(
\begin{array}{ccc}
 2a & 0 & 2c \\
 0 & 0 & 2a-c \\
 2c & 2a-c & 4c
\end{array}
\right),
\end{equation}
and
\begin{equation}
M^{\text{B}_4} =\left(
\begin{array}{ccc}
 2a &2b & 0 \\
 2b & 4b & 2a-b \\
 0 & 2a-b & 0
\end{array}
\right).
\end{equation}
We will show that all these mass matrices of type class B giving TM$_1$ mixing are not allowed
by the experimental data.

The mass matrix with TM$_1$ mixing for the class C is
\begin{equation}
M^C =\left(
\begin{array}{ccc}
 2a & 2b & 2b \\
 2b & 0 & 2a+2b \\
 2b & 2a+2b & 0
\end{array}
\right).
\end{equation}
This mass matrix has $\mu-\tau$ symmetry and implies $\theta_{13}=0$.
Hence, it is not allowed.

The phenomenology of patterns A$_{1}$ and A$_{2}$ is related:
one can obtain the predictions for A$_{2}$ by making the transformations \cite{tz,xingtz}
\begin{equation}\label{eq:trans}
\theta_{23} \rightarrow \frac{\pi}{2}-\theta_{23}, \delta
\rightarrow \pi-\delta
\end{equation}
on the predictions of A$_{1}$. Hence, we study the 
phenomenological implications for pattern A$_1$ only. 

Any mass matrix $M$ giving TM$_1$ mixing can be diagonalized by 
a mixing matrix $U = U_{\text{TM}_1}$ given in Eq. (\ref{eq:tm1})
using the relation
\begin{equation}
U^{T}MU=M_{\text{diag}}
\end{equation}
where $M_{\text{diag}}$ is the diagonal mass matrix given as 
\begin{equation}\label{eq:diag}
M_{\text{diag}} =\left(
\begin{array}{ccc}
 m_{1} & 0 & 0 \\
 0 & e^{2 i \alpha } m_{2} & 0 \\
 0 & 0 & e^{2 i \beta } m_{3}
\end{array}
\right).
\end{equation}
Here, $m_1$, $m_2$, and $m_3$ are the neutrino masses and $\alpha$
and $\beta$ are the two Majorana phases.

Once the mixing matrix $U$ is known, the mixing angles can be calculated using the relations:
\begin{equation}
  s_{12}^{2}=\frac{|U_{12}|^{2}}{1-|U_{13}|^{2}}, s_{23}^{2}=\frac{|U_{23}|^{2}}{1-|U_{13}|^{2}}
  \textrm{ and } s_{13}^{2}=|U_{13}|^{2},
\end{equation}
where $s_{ij}=\sin \theta_{ij}$ and $c_{ij}=\cos \theta_{ij}$.
For our case $U=U_{\text{TM}_1}$, the above relations give
\begin{equation}\label{eq:th12}
s_{12}^{2} = 1-\frac{2}{3-\sin^2\theta},
\end{equation}
\begin{equation}\label{eq:th23}
s_{23}^{2}=\frac{1}{2} \left(1+\frac{\sqrt{6}   \sin 2 \theta \cos\phi}{3-\sin^2\theta}\right),
\end{equation}
and
\begin{equation}\label{eq:th13}
s_{13}^{2}=\frac{1}{3}\sin^2\theta.
\end{equation}

We see from Eq. (17) that $\theta_{12}$
is smaller than its TBM value $s_{12}^{2}=1/3$. In contrast, the value of
$\theta_{12}$ is larger than the TBM value for TM$_2$ mixing. Since the
experimental value of $\theta_{12}$ is towards the lower side of the TBM value,
TM$_1$ mixing is more appealing than TM$_2$ mixing.

The CP violating phase $\delta$ can be calculated from the Jarlskog rephasing
invariant measure of CP violation \cite{jarlskog}
\begin{equation}\label{eq:jcp}
J=Im(U_{11}U^*_{12}U^*_{21}U_{22})
\end{equation}  
using the relation
\begin{equation}\label{eq:jpar}
J=s_{12}s_{23}s_{13}c_{12}c_{23}c_{13}^2 \sin \delta.
\end{equation}
Substituting the elements of the TM$_1$ mixing matrix in Eq. (\ref{eq:jcp}),
we obtain
\begin{equation}\label{eq:jtm}
J=\frac{1}{6\sqrt{6}}\sin 2 \theta \sin \phi.
\end{equation}
From Eqs. (\ref{eq:jpar}) and (\ref{eq:jtm}), we get
\begin{equation}\label{eq:delta}
\cot^2 \delta = \cot^2 \phi -\frac{6  
   \sin ^2 2 \theta \cot ^2\phi}{(3-\sin^2\theta)^2}.
\end{equation}

We reconstruct the neutrino mass matrix for TM$_1$ mixing using the relation: 
\begin{equation}
M_{\nu}=U^*M_{\text{diag}}U^{\dagger}
\end{equation}
where $U=U_{\text{TM}_1}$.
To obtain the predictions for the neutrino mass matrix 
of the type A$_1$ given by Eq. (\ref{eq:a1}), we have to solve 
the two complex equations: $M_{\nu_{11}}=0$ and $M_{\nu_{12}}=0$. 
Solving the equation $M_{\nu_{11}}=0$, we get
\begin{equation}\label{eq:r12}
\frac{m_1}{m_2}= \frac{\sin 2 (\alpha -\beta )\cos^2\theta}{2 \sin2\beta} 
\end{equation}
and
\begin{equation}\label{eq:r23}
\frac{m_2}{m_3}= - \frac{\sin2 \beta \tan ^2\theta}{\sin 2 \alpha}. 
\end{equation}
Using these two equations, we evaluate
$m_1/m_3$ and invert the resulting relation to 
obtain 
\begin{equation}\label{eq:alpha}
\cot 2\alpha=\cot 2 \beta +\frac{m_1 }{m_3}\csc 2 \beta  \cot^2 \theta.
\end{equation}
We note that the presence of a zero at (1,1)
entry in a mass matrix with TM$_1$ mixing, through Eqs. 
(\ref{eq:r12}) and (\ref{eq:r23}), implies a beautiful 
sum-rule on neutrino masses:
\begin{equation}\label{eq:sumrule}
\frac{\sin 2 (\alpha -\beta
   )}{2 m_1}
   -\frac{ \sin 2 \beta
   }{m_2}+\frac{\sin 2 \alpha
   }{m_3}=0.
\end{equation}
The texture zero at (1,1) entry in a mass matrix with TM$_1$ mixing
also gives a prediction for the ratio 
$r=\Delta m^2_{21}/\Delta m^2_{31}$. 
From Eqs. (\ref{eq:r12}) and (\ref{eq:r23}), we obtain
\begin{equation}\label{eq:rnu}
  r = \frac{- \sin^2 2( \alpha - \beta)+ 4 \sec^4 \theta \sin^2 2 \beta}
  {- \sin^2 2( \alpha - \beta)+ 4 \csc^4 \theta \sin^2 2 \beta}.
\end{equation}

We solve the second equation $M_{\nu_{12}}=0$ by equating its real and
imaginary parts to zero. We eliminate $m_1$ and $m_2$ from the resulting
equations using Eqs. (\ref{eq:r12}) and (\ref{eq:r23}). Equating the
imaginary part of $M_{\nu_{12}}$ to zero, we obtain
\begin{equation}\label{eq:sth}
  \sin^2 \theta =-\frac{\sin 2 \alpha \sin (2 \beta -\phi)}
 {\sin 2 (\alpha -\beta )\sin \phi} .
\end{equation}
We get a quadratic equation in $\tan^2\theta$ on equating the real part of
$M_{\nu_{12}}$ to zero:
\begin{eqnarray}
  \frac{\sqrt{6} \sin 2 \beta \cos (2 \alpha -\phi)}
  {\sin 2 (\alpha -\beta )} \tan^2\theta
  +3 \tan \theta \nonumber \\ 
  +\frac{\sqrt{6} \sin 2 \alpha  \cos (2 \beta -\phi)}{\sin 2 (\alpha -\beta )}=0.
\end{eqnarray}
Solving this equation by substituting $\alpha$ from Eq. (\ref{eq:alpha}), we obtain
\begin{equation}\label{eq:tth}
\tan \theta = \sqrt{\frac{3}{2}} \frac{\sin (2\beta-\phi)}{\sin 2\beta}.  
\end{equation}
The value of $\theta$ calculated in Eqs. (\ref{eq:sth}) and (\ref{eq:tth}) must be identical. This requirement gives
\begin{equation}\label{eq:alph}
\cot 2 \alpha = \cot \phi + \frac{2 \sin 2 \beta}{3 \sin (2 \beta - \phi) \sin \phi}.
\end{equation}
Equations (\ref{eq:r12}), (\ref{eq:r23}), (\ref{eq:tth}) and (\ref{eq:alph}) are the four predictions for the neutrino mass
matrix of type A$_1$. We can express these four predictions as expressions for $\alpha$, $\beta$, $\frac{m_1}{m_2}$,
$\frac{m_2}{m_3}$ as functions of $\theta$ and $\phi$. Inversion of Eq. (\ref{eq:tth}) gives 
\begin{equation}\label{eq:bet}
\cot 2 \beta = \cot \phi - \sqrt{\frac{2}{3}} \csc \phi \tan \theta.
\end{equation} 
Substituting $\beta$ from Eq. (\ref{eq:bet}) in Eq. (\ref{eq:alph}) gives 
\begin{equation}
\cot 2 \alpha = \cot \phi + \sqrt{\frac{2}{3}} \csc \phi \cot \theta.
\end{equation} 
Equations(\ref{eq:r12}) and (\ref{eq:r23}) after substituting values of $\alpha$ and $\beta$ give
\begin{equation}
\frac{m_2^2}{m_1^2} = 5 \sec^2 \theta + 4 \sqrt{6} \tan \theta \cos \phi +\tan^2 \theta - 1
\end{equation}
and
\begin{equation}
\frac{m_3^2}{m_1^2} = 6 \csc^2 \theta - 4 \sqrt{6} \cot \theta \cos \phi - 2.
\end{equation}
We can use these ratios to calculate $r = \frac{\Delta m_{21}^2}{\Delta
  m_{31}^2}$ as a function of $\theta$ and $\phi$ or we can calculate it
directly
from Eq. (\ref{eq:rnu}). We obtain 
\begin{equation}\label{eq:r}
  r = \frac{3 - 6 \sec^2 \theta - 4 \sqrt{6} \tan \theta \cos \phi}{3 -
    \csc \theta +4 \sqrt{6} \cot \theta \cos \phi}.
\end{equation}
For TM$_2$ mixing, we have $r=\tan^2\theta$ \cite{2tztm}. In contrast, $r$ is
function of both $\theta$ and $\phi$ for TM$_1$ mixing. By demanding that $r$
satisfies its experimental value, one can calculate experimentally allowed values
of $\phi$. 

Once the experimentally allowed values of $\theta$ and $\phi$ are known, the
observables $\theta_{12}, \theta_{23}, \theta_{13}$ and $\delta$ can be
calculated from Eqs. (\ref{eq:th12}), (\ref{eq:th23}), (\ref{eq:th13}) and (\ref{eq:delta}) as they are functions of $\theta$
and $\phi$.

\begin{figure*}[t]
\centering 
\includegraphics[scale=0.6]{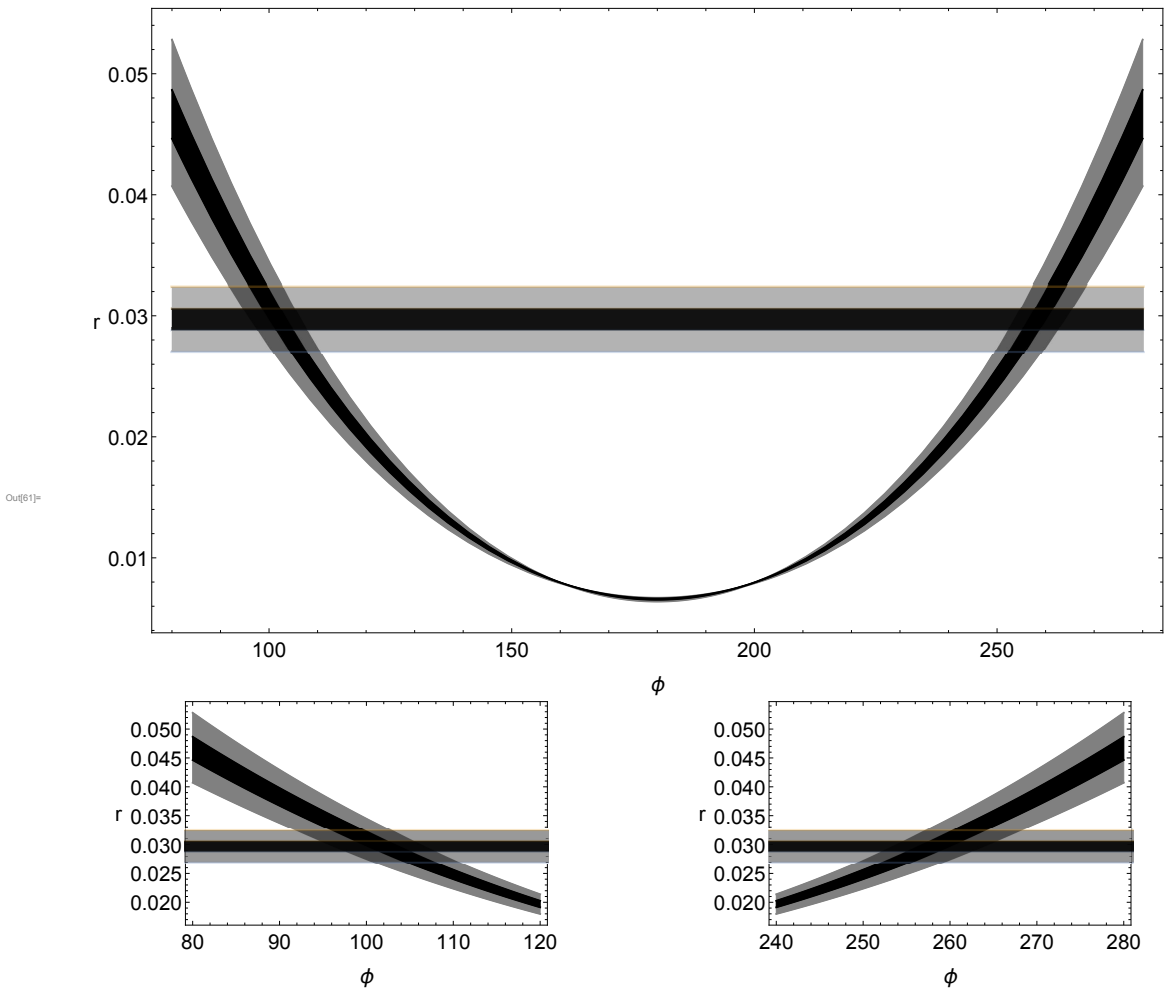}
\caption{The ratio $r=\Delta m^2_{12}/\Delta m^2_{13}$ as a function of $\phi$ (degrees).
  The horizontal line depicts the experimental value of $r$. The inner solid
  bands depict the 1$\sigma$ range and the outer gray bands depict the 3$\sigma$ range.}
\label{fig:r}
\end{figure*}

\begin{figure*}[t]
\centering 
{\includegraphics[scale=0.6]{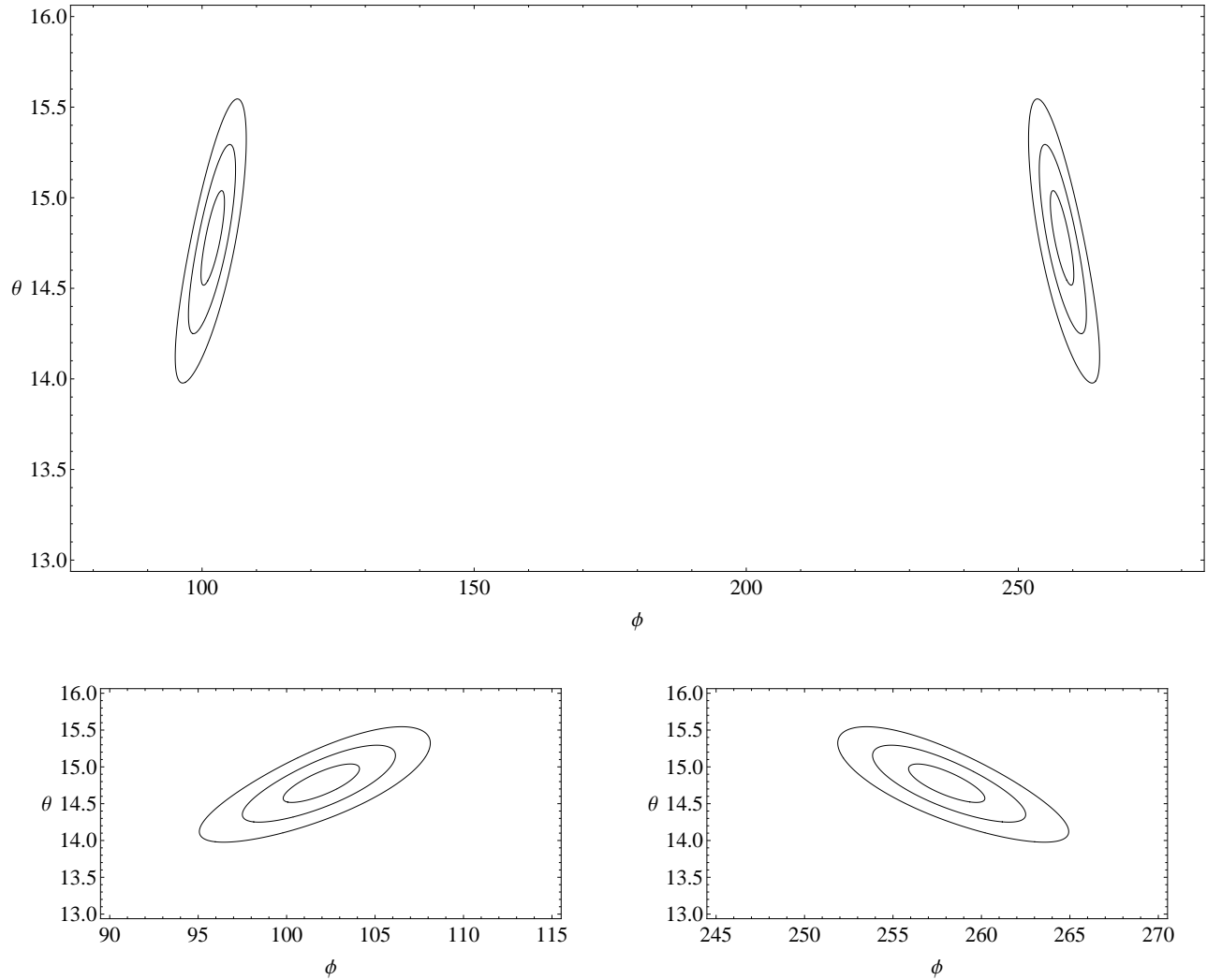}} 
\caption{The allowed region on $(\theta-\phi)$ parameter space. The contours are
at 1, 2, and 3 $\sigma$ confidence levels (CL) and all the angles are in degrees.}
\label{fig:allowed}
\end{figure*}

\begin{figure*}[t]
\centering 
\includegraphics[scale=0.7]{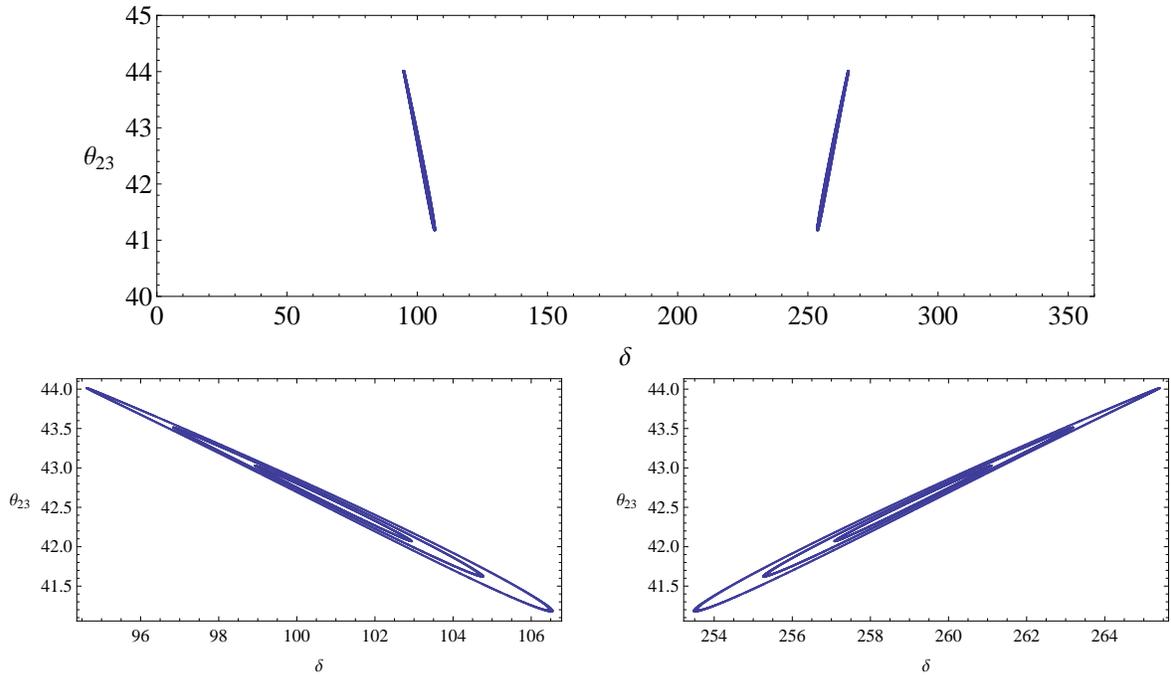}
\caption{The predictions for $\theta_{23}$ and $\delta$ at 1, 2, and 3 $\sigma$ confidence levels where all the angles are in degrees.}
\label{fig:predictions}
\end{figure*}

The experimentally allowed value of $\theta$ can be calculated from Eq. (\ref{eq:th13}) using the experimental value of $\sin^2 \theta_{13} = 0.0216 \pm 0.00075$ \cite{data}. We get $\theta = 14.78 \pm 0.26$ degrees. We could have also used the experimental value of $\sin^2 \theta_{12} = 0.306 \pm 0.012$ \cite{data} to constrain $\theta$. However, this gives a larger range of $\theta$.  We plot $r$ as a function of $\phi$ using Eq. (\ref{eq:r}) for $\theta = 14.78 \pm 0.26$ degrees in Fig. \ref{fig:r}. From the experimental values $\Delta m_{21}^2 = (7.50 \pm 0.19)\times10^{-5}$ eV$^2$ and $\Delta m_{31}^2 = (2.524 \pm 0.04)\times 10^{-3}$ eV$^2$ \cite{data}, we get the experimental value $r = 0.0297 \pm 0.0003$. Here, all the experimental errors are at one standard deviation. We find that the predicted and experimental values of $r$ are consistent only for two regions of $\phi$ depicted in Fig. \ref{fig:r}. In this way, we can constrain both $\theta$ and $\phi$. Then, all other observables can be obtained as they are functions of $\theta$ and $\phi$.\\
A better approach is to constrain $\theta$ and $\phi$ by minimizing the $\chi^2$-function
\begin{equation}
\chi^2(\theta,\phi) = \sum_{i = 1}^3 \left[\frac{v_i(\theta,\phi) - v_i^{exp}}{\sigma_i^{exp}}\right]^2
\end{equation}
where the variables $v_i = (\sin^2 \theta_{12}, \sin^2 \theta_{13}, r)$;
$v_i^{exp}$ are
experimental values of $v_i$; and $\sigma_i^{exp}$ are the standard deviations in
the experimental values of $v_i$. The $\chi^2(\theta,\phi)$ is minimum at
$\theta = 14.8$ degrees and $\phi = 103.2$ degrees or $\phi = 256.8$ degrees.
The minimum value is $\chi^2_{min}= 1.8$. The contours of $\Delta \chi^2 =
\chi^2 - \chi_{min}$ corresponding to 1$\sigma$, 2$\sigma$, and 3$\sigma$
confidence level are shown in Fig. \ref{fig:allowed}. The predictions for $\theta_{23}$ and $\delta$ for TM$_1$ mixing in pattern A$_1$ have been shown in Fig. \ref{fig:predictions} at various confidence levels.
The predictions for TM$_1$ mixing in pattern A$_2$ can be obtained by using the transformations given in Eq. (\ref{eq:trans}). We find that $\delta$ is predicted either around 100$^\circ$ or around 260$^\circ$ with a spread of around 10$^\circ$. Table \ref{tab:ranges} shows the 3$\sigma$ ranges of $\delta$ and $\theta_{23}$ for these two solutions obtained from our analysis. The mixing angle $\theta_{23}$ lies below (above) 45$^\circ$ for pattern A$_1$ (A$_2$).

Recently, long baseline neutrino oscillation experiments like MINOS and T2K  \cite{lbl} are showing a preference for the CP violating phase $\delta$ to be around 270$^\circ$. In particular, a recent global analysis in Ref. \cite{data} rules out $\delta$ from about 55$^\circ$ to 120$^\circ$ at 3$\sigma$ CL for inverted mass ordering. A certain portion of $\delta$ is also ruled out if $\theta_{23} <$ 45$^\circ$ for the normal mass ordering (see Fig. 11 in Ref. \cite{data}).  In our $\chi^2$ analysis, we do not put any experimental constraints on $\delta$ and $\theta_{23}$ as these parameters are not precisely measured and their distributions are not Gaussian. If we take into consideration the limits on $\delta$ as given in Ref. \cite{data}, the Solution - I in Table \ref{tab:ranges} for pattern A$_1$ is ruled out and only Solution - II, where $\delta$ lies around 260$^\circ$, remains compatible. Since the exclusion region of $\delta$ with respect to $\theta_{23}$ for normal mass spectrum (Fig. 11 of Ref. \cite{data}) is not symmetric around $\theta_{23} = $ 45$^\circ$, both solutions are still allowed for pattern A$_2$.

\begin{figure*}[t]
\centering 
\includegraphics[scale=0.7]{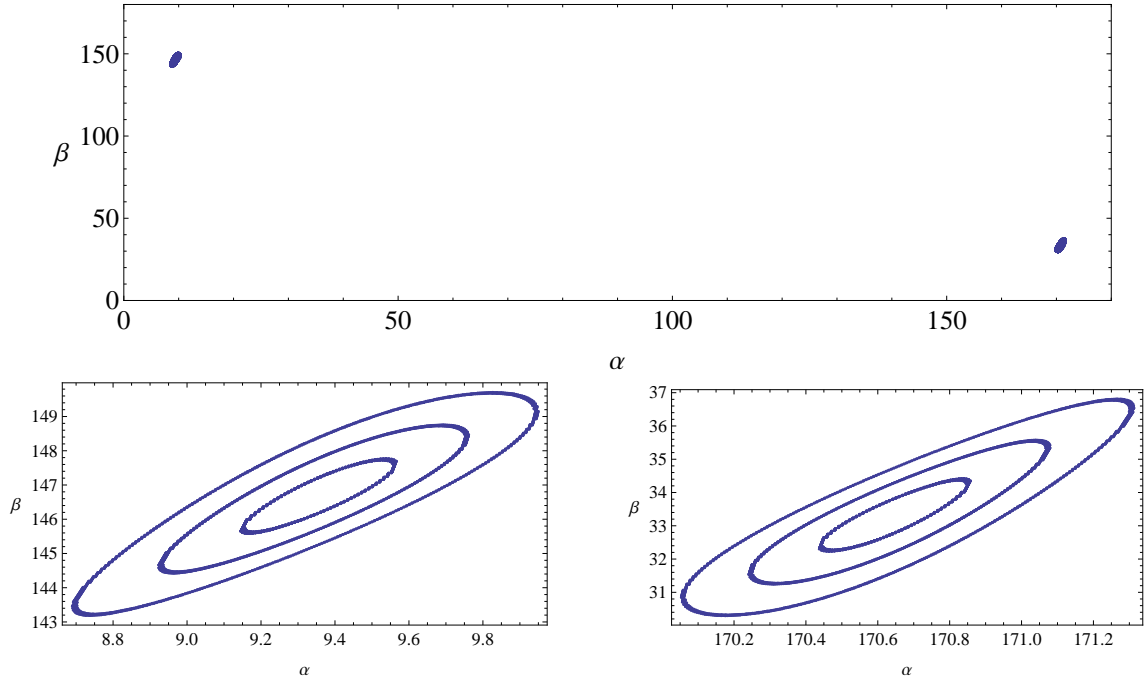}
\caption{Predictions for Majorana phases $\alpha$ and $\beta$ at 1, 2, and 3 $\sigma$ confidence levels where all the angles are in degrees.}
\label{fig:ab}
\end{figure*}

\begin{figure*}[t]
\centering 
\includegraphics[scale=0.7]{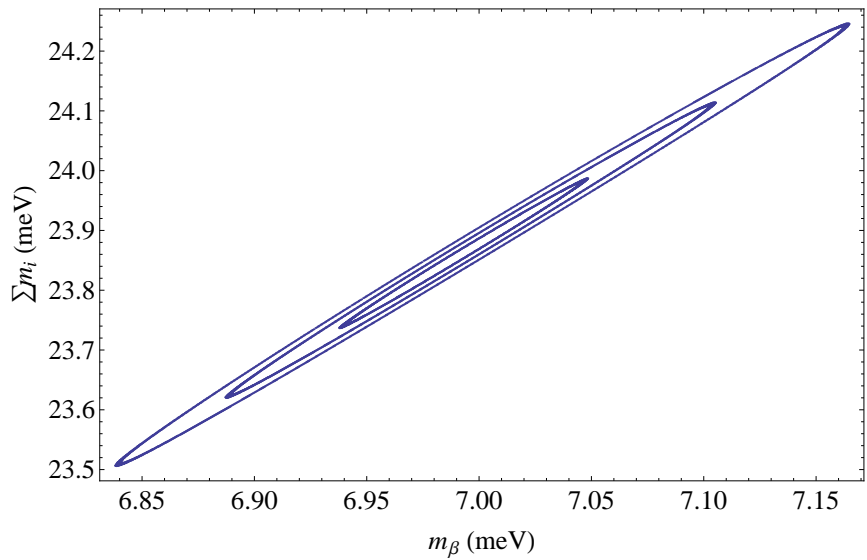}
\caption{Predictions for effective electron neutrino mass for $\beta$-decay and sum of neutrino masses at 1, 2, and 3 $\sigma$ confidence levels.}
\label{fig:sum}
\end{figure*}

The predictions for Majorana phases $\alpha$ and $\beta$ are shown in Fig. \ref{fig:ab}. We also depict the predictions for the effective electron neutrino mass for $\beta$-decay \mbox{$m_\beta^2 = m_1^2 |U_{e1}|^2 + m_2^2 |U_{e2}|^2 + m_3^2 |U_{e3}|^2$} and the sum of neutrino masses \mbox{$\sum m_i = m_1 + m_2 + m_3$} in \mbox{Fig. \ref{fig:sum}}. Since the (1,1) element of the neutrino mass matrix vanishes for patterns A$_1$ and A$_2$, this leads to a vanishing effective Majorana neutrino mass ($|M_{ee}| = |m_1 U_{e1}^2 + m_2 U_{e2}^2 + m_3 U_{e3}^2|$) for these patterns.

\begin{table}[tb]
\begin{center}
\begin{tabular}{ccccc}
\hline
\hline
\multirow{2}{*}{Type} & \multirow{2}{*}{$\theta_{23}$}& \multicolumn{2}{c}{$\delta$}\\
  &    & Solution - I &  Solution - II &  \\
\hline
  A$_1$ &41.18$^\circ$ - 44.02$^\circ$ & 94.6$^\circ$ - 106.5$^\circ$ & 253.5$^\circ$ - 265.5$^\circ$ \\
 A$_2$ &45.98$^\circ$ - 48.82$^\circ$& 73.5$^\circ$ - 85.5$^\circ$& 274.6$^\circ$ - 286.5$^\circ$ \\
 \hline
 \hline
\end{tabular}
\end{center}
\caption{Allowed 3$\sigma$ ranges of $\theta_{23}$ and $\delta$ for patterns A$_1$ and A$_2$. Solution - I and Solution - II are two allowed solutions for $\delta$.}
\label{tab:ranges}
\end{table}

\begin{figure*}[t]
\centering 
\includegraphics[scale=0.8]{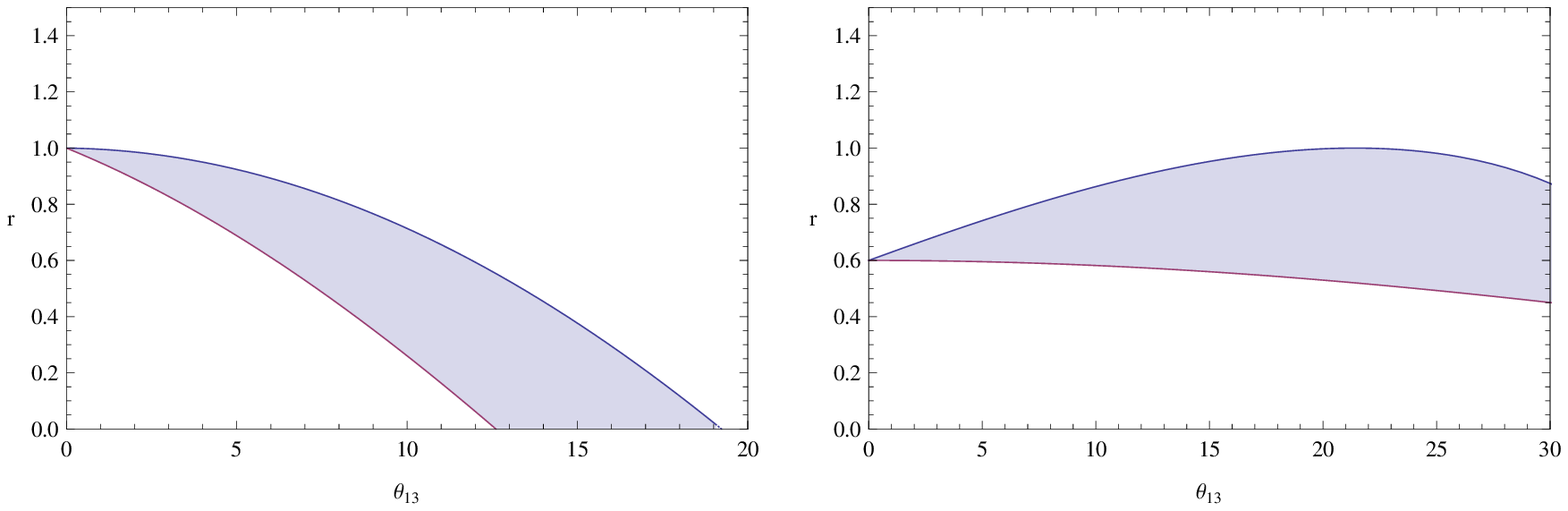}
\caption{The inconsistency between $r$ and $\theta_{13}$ (degrees) for patterns B$_1$ (left) and B$_3$ (right).}
\label{fig:b1b3}
\end{figure*} 

The neutrino mass matrix of type B$_1$ has zeros at $(1,3)$ and $(2,2)$
entries. This implies following expression of ratio $r$ in the presence of TM$_1$ mixing:
\begin{equation}
r=\frac{1}{6} \left(2 \sqrt{6} \sin (2 \theta) \cos (\phi)+9 \cos (2 \theta)-3\right).
\end{equation}
We can express $\theta$ in terms of $\theta_{13}$ using 
Eq. (\ref{eq:th13}) in the above relation expressing $r$ as a
function of $\theta$ \mbox{(Fig. \ref{fig:b1b3})}.
It is clear that we cannot have both $r$ and $\theta_{13}$
in their experimentally allowed ranges simultaneously. Hence, this pattern
is inconsistent with the experimental data when combined with TM$_1$ mixing. 
The neutrino mass matrix of type B$_2$ is related to the neutrino mass matrix
of type B$_1$ by a $\mu$-$\tau$ exchange \cite{tz,xingtz} and has identical predictions for
$r$ and $\theta_{13}$. Hence, neutrino mass matrix of type B$_2$ with TM$_1$
mixing is also incompatible with the recent experimental data.

The neutrino mass matrix of type B$_3$ has zeros at $(1,2)$ and $(2,2)$
entries. The expression of ratio $r$ in the presence of TM$_1$ mixing is given by
\begin{equation}
r=\frac{1}{10} \left(-2 \sqrt{6} \sin (2 \theta) \cos (\phi)-\cos (2 \theta)-5\right).
\end{equation} 
In case of pattern B$_3$, the parameter $r$ always remains larger than 0.4 whereas the experimental range of this parameter lies well below the value 0.4 (see \mbox{Fig. \ref{fig:b1b3}}). Thus pattern B$_3$ is also incompatible with the experimental data. Since pattern B$_3$ is related to pattern B$_4$ by $\mu$-$\tau$ exchange symmetry, the pattern B$_4$ is also incompatible with the experimental data when combined with TM$_1$ mixing.

In conclusion, we have studied the phenomenological implications of two texture zeros in the presence of TM$_1$ mixing. 
There are seven allowed patterns for the presence of two texture zeros in the neutrino mass matrix. The presence of TM$_1$ mixing rules out five out of the seven patterns of two texture zeros. The neutrino mass matrix having two texture zeros and TM$_1$ mixing simultaneously can only belong to patterns A$_1$ and A$_2$. The Dirac CP violating phase $\delta$ is restricted to two narrow regions around $100^{\circ}$ and  $260^{\circ}$ for these patterns. For TM$_1$ mixing, $\theta_{12}$ is smaller than its TBM value and moves towards its best fit value with the increase in $\theta$. The imposition of TM$_1$ mixing on two zeros make these classes very predictive and these predictions can be tested in future neutrino oscillation experiments. 

\acknowledgements{R. R. G. acknowledges the  financial support provided by Department of Science and Technology, Government of India under the Grant No. SB/FTP/PS-128/2013.}

\end{document}